\documentclass[12pt]{article}
\usepackage{graphicx,amsmath,amssymb,stackrel,ytableau,simplewick,color,url}
\usepackage[a4paper, left=2.3cm, right=2.3cm, top=2.3cm, bottom=2.3cm]{geometry}
\usepackage{hyperref}
\usepackage{multirow}
\hypersetup{
    colorlinks=true,
    linkcolor=blue,
    filecolor=magenta,      
    urlcolor=blue,
    citecolor=purple
}
\newcommand{\beq}{\begin{equation}}
\newcommand{\eeq}{\end{equation}}
\newcommand{\beqs}{\begin{eqnarray}}
\newcommand{\eeqs}{\end{eqnarray}}
\newcommand{\bit}{\begin{itemize}}
\newcommand{\eit}{\end{itemize}}
\newcommand{\bce}{\begin{center}}
\newcommand{\ece}{\end{center}}
\newcommand{\ben}{\begin{enumerate}}
\newcommand{\een}{\end{enumerate}}

\newcommand{\hc}{\mathrm{h.c.}}
\newcommand{\diag}{\mathrm{diag}}
\newcommand{\nn}{\nonumber}

\begin{document}
\pagestyle{empty}

\begin{center}

{\LARGE{\bf Composite ALPs in Composite Higgs Models}}

\vspace{1.8cm}

{\large{Gabriele Ferretti}}

\vspace{1.2cm}

{\it Department of Physics, \\
Chalmers University of Technology, \\
Fysikg{\aa}rden 1, 41296 G\"oteborg, Sweden\\
{\tt ferretti@chalmers.se}}

\vspace{2.cm}

\begin{minipage}[h]{14.0cm}

\bce {\bf \large Abstract} \ece

\medskip
We study the effective Lagrangian for the Axion-Like Particle (ALP) arising in a class of composite Higgs models. We discuss the peculiarities of the coupling to the quark sector specific of these models, including flavor violating terms. We then confront the main bounds from rare $B$-meson decays.

\end{minipage}
\end{center}
\newpage

%  Resetting of counters
\setcounter{page}{1} \pagestyle{plain} \renewcommand{\thefootnote}{\arabic{footnote}} \setcounter{footnote}{0}

\section{Introduction}
\label{sec:intro}

There is a class of Composite Higgs Models (CHM) predicting the existence of an Axion-Like Particle (ALP) $a$ with a mass below the Electro-Weak (EW) scale. These are models based on underlying four-dimensional gauge theories with hyperfermions, where the Higgs boson arises as a pseudo-Nambu-Goldstone boson (pNGB) as in \cite{Kaplan:1983sm}, and top partners arise as fermionic composite states as in Partial Compositeness (PC) \cite{Kaplan:1991dc}.

Models of this type have been proposed and classified in \cite{Barnard:2013zea,Ferretti:2013kya}\footnote{The classification in \cite{Ferretti:2013kya} included both conformal and confining models. A streamlined presentation is given in \cite{Ferretti:2016upr}.}. The presence of an ALP is easily understood by noticing that these models employ two sets of hyperfermions $\psi$ and $\chi$, transforming under a different representation of the hypercolor gauge group. One linear combination of their two $U(1)$ axial symmetries is thus free of ABJ anomalies with the hypercolor group. Its breaking gives rise to a light pseudoscalar whose mass is generated by explicit symmetry breaking terms such as hyperquark masses or couplings with the Standard Model (SM) fields.

The most realistic scenario is one where the ALP $a$ is the only BSM particle left below the EW scale.
A simplified Lagrangian was used for phenomenological studies in \cite{Cacciapaglia:2017iws,Cacciapaglia:2019bqz,Borsato:2021aum,BuarqueFranzosi:2021kky} where the couplings to the SM fermions were taken to be diagonal in the mass basis and degenerate except for the top quark. 
This assumptions needs to be revisited in the light of the extensive study on ALP Lagrangians done in \cite{Bauer:2017ris,Bauer:2020jbp,Bauer:2021mvw}, building on the Lagrangian in \cite{Georgi:1986df}, and this is the goal of this note. 

We present a more general version of the ALP Lagrangian in CHM that takes into account the Renormalization Group (RG) evolution of the couplings. Crucially, unavoidable Flavor Violating (FV) couplings arise from the evolution from the UV scale. We present the relevant formulas and confront them with the experimental limits for this specific incarnation of the ALP.

In its full generality the effective Lagrangian depends on five dimensionless couplings to the quarks (six minus an overall shift), one dimensionless coupling to the leptons (two minus an overall shift) and three dimensionless couplings to the SM vector bosons. The dimensionfull parameters are the ALP mass $m_a$, and its decay constant $f$ that can always be set to 1~TeV by a rescaling of the dimensionless constants.

The paper is organized as follows.
In Sec.~\ref{sec:motivating} we motivate the structure of the Lagrangian by discussing the peculiarities of the ALP couplings to the quarks specific of CHMs. In Sec.~\ref{sec:ALPlag} we present the full Lagrangian, in its two equivalent forms, valid at the UV scale. In Sec.~\ref{sec:running} we discuss the running of the coupling to the EW scale and below.
Finally, in Sec.~\ref{sec:ALPpheno} we discuss the phenomenology arising from this Lagrangian, confront the bounds from rare $B$ decay, and discuss the applicability of the simplified Lagrangian previously used for collider studies.

\section{Motivating the Lagrangian}
\label{sec:motivating}

The purpose of this section is to motivate the general form of the Lagrangian to be presented in Sec.~\ref{sec:ALPlag}.  The main difference from other ALP Lagrangians arises in the coupling to the quarks. Hence, in this section, we focus on the quark sector and \emph{we ignore} the leptons, (where no FV occurs), and the gauge bosons, (whose couplings to the ALP are affected by the anomaly in a known way).

A generic $\dim=5$ interaction is of the type
\begin{equation}
    \mathcal{L}_{\mathit{gen.}} =
    - \frac{a}{f}\left( \bar{Q}_L\tilde{Y}_u\tilde{\Phi} u_R + \bar{Q}_L\tilde{Y}_d \Phi d_R \right) +\hc
\label{Lchiralbasis}
\end{equation}
where $\Phi$ is the Higgs doublet, $Q_L,u_R,d_R$ are the left/right handed $SU(2)_L$ doublets/singlets with three generations, and $\tilde{Y}_u, \tilde{Y}_d$ are $3\times 3$ complex matrices, for a total of 36 real parameters. For generic $\tilde{Y}_u, \tilde{Y}_d$ in (\ref{Lchiralbasis}) the Lagrangian is not shift invariant and does not describe an ALP. In order to be able to implement shift invariance, the $\tilde{Y}_u, \tilde{Y}_d$ must be related to the SM Yukawa couplings $Y_u, Y_d$ by the relations
\begin{equation}
      \tilde{Y}_u = i(Y_u C_u - C_Q Y_u), \quad  \tilde{Y}_d = i(Y_d C_d - C_Q Y_d), \label{shifty}
\end{equation}
with $C_u, C_d, C_Q$ $3\times 3$ hermitian matrices\footnote{We denote the various ALP flavor matrices by capital $C_{\dots}$ and their elements by lower $c_{\dots}$.}, for a total of 27 real parameters.

A way to show (\ref{shifty}) is to note that, if we take the part of the SM Lagrangian containing the quark fields
\begin{equation}
    {\mathcal{L}}_{\mathit{q SM}} = \bar{Q}_L i  \!\! \not \!\!  D Q_L +\bar{u}_R i  \!\! \not \!\!  D u_R +\bar{d}_R i  \!\! \not \!\!  D  d_R - \left(\bar{Q}_L Y_u\tilde{\Phi} u_R + \bar{Q}_L Y_d \Phi d_R  + \hc\right), \label{lagrangianSM}
\end{equation}
and couple the ALP by adding a (manifestly shift invariant) derivative interaction
\begin{equation}
     {\mathcal{L}}_{\partial a} =\frac{\partial_\mu a}{f}\left( \bar{Q}_L C_Q \gamma^\mu  Q_L +  \bar{u}_R C_u \gamma^\mu  u_R +\bar{d}_R C_d \gamma^\mu  d_R \right), \label{lagrangianderivative}
\end{equation}
the interaction can be ``unwound'' by performing an ALP dependent flavor rotation
\begin{equation}
    Q_L\to e^{i C_Q \frac{a}{f}} Q_L, \quad u_R\to e^{i C_u \frac{a}{f}} u_R, \quad d_R\to e^{i C_d \frac{a}{f}} d_R,
\end{equation}
Leading to
\begin{align}
    &{\mathcal{L}}_{\mathit{q SM}}+{\mathcal{L}}_{\partial a}\to \label{Lshiftspurion} \\& \bar{Q}_L i \!\! \not \!\!  D  Q_L +\bar{u}_R i  \!\! \not \!\!  D  u_R +\bar{d}_R i \!\! \not \!\!  D d_R - \left(\bar{Q}_L  e^{-i C_Q \frac{a}{f}} Y_u\tilde{\Phi} e^{i C_u \frac{a}{f}} u_R + \bar{Q}_L  e^{-i C_Q \frac{a}{f}} Y_d \Phi e^{i C_d \frac{a}{f}} d_R  + \hc\right). \nn
\end{align}
To first order in $1/f$, the Lagrangian (\ref{Lshiftspurion}) reduces to (\ref{lagrangianSM}) plus (\ref{Lchiralbasis}), with the choice (\ref{shifty}).

It is the Lagrangian (\ref{Lshiftspurion}) that is in the typical form one gets when doing a spurion analysis of a BSM theory. One assigns various charges to the Yukava couplings and writes down the formally invariant terms. 
However, in CHM, the spurion Lagrangian is not quite of the form (\ref{Lshiftspurion}). Instead, a generic situation is where the interaction Lagrangian takes the form
\begin{align}
    {\mathcal{L}}_{\mathit{CHM}} = -\bigg(&e^{i q_u^{(1)}\frac{a}{f}}\bar{Q}_LY^{(1)}_u\tilde{\Phi} u_R + e^{i q_u^{(2)}\frac{a}{f}}\bar{Q}_LY^{(2)}_u\tilde{\Phi} u_R \nn\\ 
    +  & e^{i q_d^{(1)}\frac{a}{f}}\bar{Q}_LY^{(1)}_d \Phi d_R + e^{i q_d^{(2)}\frac{a}{f}}\bar{Q}_LY^{(2)}_d \Phi d_R + \hc\bigg), \label{lagrangianPC}
\end{align}
i.e. it is the sum of two terms with formal $U(1)$ charges $q^{(i)}$ for each sector, one coming from the bilinear sector, \`a la Technicolor, and one from the PC linear sector.

The origin of this structure of the couplings, limited to the top quark, is discussed in \cite{Belyaev:2016ftv}. The general case can be exemplified as follows. Recall from the Introduction that these models are based on a gauge theory with two sets of hyperfermions $\psi$ and $\chi$. We assign to them $(U(1)_\psi,U(1)_\chi)$ axial charges $(1,0)$, and $(0,1)$ respectively. 

We can formally maintain these symmetries by assigning various charges to the couplings to the SM fields thought of as spurions. For example, a bilinear Yukawa coupling $y$ of the type\footnote{We briefly revert to Weyl notation in this paragraph.} $y Q_L \psi\psi u_R^c$ will carry charges $\mathbf{n}_y = (-2,0)$, while PC pre-Yukawa couplings of the type, say, $g Q_L \chi\psi\chi$ and $g' u_R^c\chi^\dagger\psi^\dagger \chi$ will carry charges $\mathbf{n}_g = (-1,-2)$ and $\mathbf{n}_{g'} = (1,0)$, leading to a Yukawa coupling $gg'$ of charges $\mathbf{n}_{gg'} = \mathbf{n}_g + \mathbf{n}_{g'} = (0,-2)$. There are a discrete number of possibilities for these charge assignments \cite{Belyaev:2016ftv}.

Eq. (\ref{lagrangianPC}) denotes the general expression one obtains when two different combinations of couplings are involved. The $U(1)$ charges $q^{(i)}$ are the projections of the above Yukawa charges $\mathbf{n}_i$ along the hypercolor anomaly-free\footnote{The condition on $q_\psi, q_\chi$ for the absence of ABJ anomalies with the hypercolor group is $q_\psi N_\psi I_\psi + q_\chi N_\chi I_\chi = 0$, where $N_{\psi}, N_{\chi}$ is the number of Weyl fermions and $I_{\psi}, I_{\chi}$ the index of their hypercolor representation.} direction $\mathbf{q}_a = (q_\psi, q_\chi)$ associated to the ALP: $q^{(i)} \propto \mathbf{q}_a\cdot \mathbf{n}_i$, for each sector $u$ and $d$. The proportionality constant can be absorbed in the definition of $f$. The physics only depends on the ratio $q_\chi/q_\psi$.

Let us now define
\begin{equation}
    q_u=\frac{q_u^{(1)}+q_u^{(2)}}{2}, \quad  \Delta q_u=\frac{q_u^{(1)}-q_u^{(2)}}{2},
    \quad Y_u = Y^{(1)}_u + Y^{(2)}_u, \quad  \quad \Delta Y_u = Y^{(1)}_u - Y^{(2)}_u, \label{deltas}
\end{equation}
and similarly for the $d$-sector\footnote{$Y_u$ and $Y_d$ are the true SM Yukawa couplings.}. Condition (\ref{shifty}) becomes
\begin{align}
    q_u Y_u + \Delta q_u \Delta Y_u &= Y_u C_u - C_Q Y_u \nn\\
    q_d Y_d + \Delta q_d \Delta Y_d &= Y_d C_d - C_Q Y_d. \label{notquite}
\end{align}
Again, this condition has no solutions generically\footnote{It gives 36 linear equations with inhomogeneous terms $\Delta Y_u, \Delta Y_d$ for the set of 27 unknown variables $C_Q, C_u, C_d$. The rank of the homogeneous system is actually 26 since one can shift $C_u \to C_u + k{\mathbf{1}}$, $C_d\to C_d + k{\mathbf{1}}$, and $C_Q\to C_Q + k{\mathbf{1}}$.}.

However, specific models of PC, where the third quark family is treated differently, lead to regions of parameter space where (\ref{notquite}) is satisfied or very close to be satisfied. To show an example, let us preliminarily notice that one can always, without breaking $SU(2)_L$ or losing  generality, chose a chiral basis where 
$Y_u=\Delta_u$ and $Y_d=V_{CKM}\Delta_d$, with $\Delta_u =\diag(y_u, y_c, y_t)$, $\Delta_d=\diag(y_d, y_s, y_b)$ diagonal matrices of the physical Yukawa couplings and $V_{CKM}$ the CKM matrix \cite{Cabibbo:1963yz,Kobayashi:1973fv}. From now on we will always work in this basis for definitiveness\footnote{The reason why one prefers to have $Y_u$ diagonal, and not $Y_d$, is that one wants eventually to integrate out the top quark.}.

With the above choice of Yukawa couplings, in the Cabibbo approximation for the CKM matrix
\begin{equation}
    V_C=
    \begin{pmatrix}
    \cos\theta_C & \sin\theta_C & 0\\
    -\sin\theta_C &  \cos\theta_C & 0\\
    0 & 0 & 1
\end{pmatrix}, \label{Cabibbo}
\end{equation}
the following matrices, aligned along the third generation
\begin{equation}
    \Delta Y_u=
    \begin{pmatrix}
    0 & 0 & 0\\
    0 & 0 & 0\\
    0 & 0 & \Delta y_t
\end{pmatrix},
\quad 
 \Delta Y_d \equiv V_C \Delta Y_d=
    \begin{pmatrix}
    0 & 0 & 0\\
    0 & 0 & 0\\
    0 & 0 & \Delta y_b
\end{pmatrix}, 
\end{equation}
admit the exact diagonal solution\footnote{Note that the first and second diagonal entries are the same, e.g. $c_{u_R}=c_{c_R}$, and we use the first flavor to label both. For $C_Q$, we use the $u$-type quark names, since we are using the chiral basis where $Y_u$ is diagonal.}
\begin{equation}
    C_Q=
    \begin{pmatrix}
    c_{u_L} & 0 & 0\\
    0 & c_{u_L} & 0\\
    0 & 0 & c_{t_L}
\end{pmatrix},
\quad 
    C_u=
    \begin{pmatrix}
    c_{u_R} & 0 & 0\\
    0 & c_{u_R} & 0\\
    0 & 0 & c_{t_R}
\end{pmatrix},
\quad     
    C_d=
    \begin{pmatrix}
    c_{d_R} & 0 & 0\\
    0 & c_{d_R} & 0\\
    0 & 0 & c_{b_R}
\end{pmatrix}, \label{assumptions}
\end{equation}
with the conditions
\begin{align}
    & q_u = c_{u_R} - c_{u_L}, \nn \\
    & q_d = c_{d_R} - c_{u_L}, \nn \\
    & q_u y_t + \Delta q_u \Delta y_t = y_t (c_{t_R} - c_{t_L}), \nn \\
    & q_d y_b + \Delta q_d \Delta y_b = y_b (c_{b_R} - c_{t_L}). \label{underdet}
\end{align}

We have gone from an over-determined system (\ref{notquite}) to an under-determined one (\ref{underdet}). In fact, (\ref{underdet}) admits a two-dimensional space of solutions spanned by $c_{u_L}$ and $c_{t_L}$. In particular, there is one subspace where $c_{u_L}=c_{t_L}$, which corresponds to absence of FV at that scale, but this condition is not preserved by the RG evolution, as we will discuss in Sec.~\ref{sec:running}.

We thus make the crucial assumption that the couplings of the ALP to the quarks in CHM are characterized by (\ref{assumptions}). 
Working backwards from eq. (\ref{notquite}) this tells us what kind of terms $\Delta Y_u$ and $\Delta Y_d$ we are allowing in a general CHM. The parameterization (\ref{assumptions}) stand out as the minimal deformation of the flavor preserving case that is also closed under the RG evolution if one ignores the contributions of the Yukawas of the first two generations \cite{Bauer:2020jbp,Bauer:2021mvw}.

The amount of FV is built into the difference $c_{u_L}-c_{t_L}$, since this is what controls $[C_Q, V_{CKM}]$ (see Sec.~\ref{sec:ALPlag}). We will see in Sec.~\ref{sec:running} that the choice (\ref{assumptions}) is preserved by RG. It is only the stronger conditions $C_Q\propto {\mathbf{1}}$, $C_u\propto {\mathbf{1}}$, $C_d\propto {\mathbf{1}}$ that are not maintained. In fact, if one were only to retain the top Yukawa $y_t$ contribution to the RG evolution, one could also set $c_{b_R}=c_{d_R}$, i.e. $C_d\propto {\mathbf{1}}$. Not doing so allows one to consider the next-to-leading situation where also $y_b\not = 0$ and the CKM matrix is approximated by (\ref{Cabibbo}).

In previous phenomenological work \cite{Cacciapaglia:2017iws,Cacciapaglia:2019bqz,Borsato:2021aum,BuarqueFranzosi:2021kky} we coupled the CHM ALP to the SM as
\begin{align}
\mathcal{L}_{\mathit{simp.}} =
- i\frac{a}{f}\sum_{\psi\not=t} c\, m_\psi\bar{\psi}\gamma^5\psi - i\frac{a}{f} c_t m_t\bar{t}\gamma^5 t +\frac{a}{4\pi f}\sum_V c_V\alpha_V V_{\mu\nu}\tilde V^{\mu\nu},
 \label{naiveLagrangian}
\end{align}
where $\psi$ are all SM fermions other than the top quark $t$, and $V$ are the SM gauge bosons with their ``fine structure constants'' $\alpha_V$. We will see that (\ref{naiveLagrangian}) can be reproduced by the highly non-generic assumption that $C_Q \propto {\mathbf{1}}$ at the EW scale. 
Even with this optimistic assumption, some small amount of Flavor Violation will be introduced by matching, after integrating out the $t, H, Z, W$. We do not dwell upon it now because we are going to analyze the more general case in Sec.~\ref{sec:running}.

\section{The ALP Lagrangian}
\label{sec:ALPlag}

We now reintroduce all SM fields ignored in Sec.~\ref{sec:motivating} and present the starting form of the ALP interaction Lagrangian,
\begin{align}
    {\mathcal{L}}_{alp} =& \frac{\partial_\mu a}{f}\left( \bar{Q}_L C_Q \gamma^\mu  Q_L +  \bar{u}_R C_u \gamma^\mu  u_R +\bar{d}_R C_d \gamma^\mu  d_R + \bar{L}_L C_L \gamma^\mu L_L + \bar{e}_R C_e \gamma^\mu  e_R\right)\nn\\
    &+ \frac{a}{4\pi f}\left(c_{GG} \alpha_s G_{\mu\nu}^a\tilde{G}^{a \mu\nu} + c_{WW} \alpha_w W_{\mu\nu}^i\tilde{W}^{i\mu\nu} + c_{BB} \alpha_Y B_{\mu\nu}\tilde{B}^{\mu\nu}\right).
\end{align}
We use standard notation for the SM fields and couplings.
We always set the redundant ALP-Higgs operator $i(\partial_\mu a/f)\Phi^\dagger D^\mu \Phi$ to zero by the usual ALP dependent field redefinition proportional to the hypercharge \cite{Georgi:1986df}.

We take $C_Q, C_u, C_d$ diagonal, as in (\ref{assumptions}), but not proportional to the identity, while $C_L = c_{e_L}{\mathbf{1}}$ and $C_e = c_{e_R}{\mathbf{1}}$, since we do not introduce PC for the leptons. The mass of the ALP is generated by masses for the hyperquarks and possible additional shift symmetry breaking terms in the UV. 

We fix $f=1\mbox{ TeV}$ and consider the couplings $c_{u_R}$, $c_{t_R}$, $c_{d_R}$, $c_{b_R}$, $c_{u_L}$, $c_{t_L}$, $c_{e_R}$, $c_{e_L}$, $c_{GG}$, $c_{WW}$, $c_{BB}$ as ``free'' parameters of order one at the scale $\Lambda=4\pi f\approx 13\mbox{ TeV}$. They can be estimated given a specific UV completion of the theory. A change in $f$ can be absorbed into a rescaling of the $c_{\dots}$. One can also use baryon and total lepton symmetry to shift the couplings to the quarks or those to the leptons by an overall constant. The total number of independent parameters is thus nine $c_{\dots}$ (5 quarks, 1 lepton, and 3 gauge), as well as the ALP mass $m_a$.

An alternative and equivalent version of the interaction Lagrangian to order $1/f$ is
\begin{align}
    {\mathcal{L}}_{alp} =& - \frac{a}{f}\left( \bar{Q}_L\tilde{Y}_u\tilde{\Phi} u_R + \bar{Q}_L\tilde{Y}_d \Phi d_R + \bar{L}_L\tilde{Y}_e \Phi e_R + \hc \right)\nn\\
    &+ \frac{a}{4\pi f}\left(\tilde c_{GG} \alpha_s G_{\mu\nu}^a\tilde{G}^{a\mu\nu} + \tilde c_{WW} \alpha_w W_{\mu\nu}^i\tilde{W}^{i\mu\nu} + \tilde c_{BB} \alpha_y B_{\mu\nu}\tilde{B}^{\mu\nu}\right),\label{alternativelagrangian}
\end{align}
with
\begin{equation}
      \tilde{Y}_u = i(Y_u C_u - C_Q Y_u), \quad  \tilde{Y}_d = i(Y_d C_d - C_Q Y_d), \quad  \tilde{Y}_e = i(Y_e C_e - C_L Y_e), \label{allshifty}
\end{equation}
and, because of the anomaly
\begin{align}
    \tilde{c}_{GG} &= c_{GG} + c_{u_R} + \frac{1}{2}c_{t_R} + c_{d_R} + \frac{1}{2}c_{b_R} - 2 c_{u_L} - c_{t_L},\nn\\
    \tilde{c}_{WW} &= c_{WW} - 3 c_{u_L} - \frac{3}{2}c_{t_L} - \frac{3}{2}c_{e_L},\nn\\
    \tilde{c}_{BB} &= c_{BB} + \frac{8}{3}c_{u_R} + \frac{4}{3}c_{t_R} + \frac{2}{3}c_{d_R} + \frac{1}{3}c_{b_R}  
         - \frac{1}{3}c_{u_L} - \frac{1}{6}c_{t_L} + 3 c_{e_R} -\frac{3}{2}c_{e_L}. \label{CVV}
\end{align}

Given a UV model, the gauge couplings can be computed from the anomaly content of the hyperfermions as shown in \cite{Belyaev:2016ftv}. Notice that the anomaly of the hyperfermions is computing directly $\tilde c_{VV}$, not $c_{VV}$. To build an intuition for this fact, recall the examples of the KSVZ axion \cite{Kim:1979if,Shifman:1979if} and the DFSZ axion \cite{Zhitnitsky:1980tq,Dine:1981rt}. 

In the KSVZ case\footnote{The composite axion \cite{Kim:1984pt} is even closer in spirit to the CHM.}, the coupling of the axion to the gluon arises entirely from the BSM sector, due to the anomalous $U(1)_{PQ}$ symmetry \cite{Peccei:1977hh,Weinberg:1977ma,Wilczek:1977pj} of the heavy BSM fermions. No derivative or Yukawa couplings to the SM fermions arise at this level. In this case the difference between $c_{GG}$ and $\tilde c_{GG}$ is irrelevant.

The DFSZ case is, in a sense, the opposite. The BSM sector is purely bosonic and no anomaly is present. The axion is a specific combination of the Higgs fields. Its coupling to the SM fermions arises from the usual Higgs Yukawa couplings. So, at this stage one has a Yukawa-type coupling of the SM fermions with the axion, and no gluon coupling ($\tilde c_{GG}=0$). When rotating the SM fermions to remove the axion from the Yukawas one introduces the axion-current derivative coupling and the axion-gluon coupling $c_{GG}\not =0$. 

The case of CHM contains both phenomena. There is a $U(1)SU(3)^2$ anomaly present in the BSM sector as in KSVZ. This leads to a direct contribution to the axion-gluon coupling from the BSM sector. At the same time, the axion also has Yukawa couplings with the fermions as in DFSZ, arising from the four fermi terms connecting the BSM and SM sectors. So we have both a $\tilde c_{GG}\not =0$, purely induced by the BSM sector, and an axion-Yukawa coupling. We can then rotate the axion as in the DFSZ case, and work with $c_{GG}\not =0$ and derivative coupling, if we wish.

The fermionic couplings are more model dependent. They can also be estimated, given some assumptions on the spurions \cite{Belyaev:2016ftv}, but it is better to think of these estimates as reasonable ``benchmarks'' and not as hard predictions from the models.

If one wants to use the ALP Lagrangian directly at the UV scale, one only needs to rotate the fields into their mass eigenbasis. Switching now to the physical Fermi fields, we obtain (dropping the coupling to the conserved currents)
\begin{align}
    {\mathcal{L}}_{alp} =&\frac{\partial_\mu a}{2f}\sum_{\psi} \xi_{\psi\psi} \bar{\psi}\gamma^\mu\gamma^5\psi+\frac{\partial_\mu a}{f}\left(\xi_{ds}\bar{d}\gamma^\mu P_L s + \xi_{db}\bar{d}\gamma^\mu P_L b + \xi_{sb}\bar{s}\gamma^\mu P_L b + \hc\right)\nn\\
    &+ \frac{a}{4\pi f}\left(c_{GG} \alpha_s G_{\mu\nu}^a\tilde{G}^{a\mu\nu} + c_{\gamma \gamma} \alpha F_{\mu\nu}\tilde{F}^{\mu\nu} + c_{\gamma Z} \frac{2\alpha}{s_w c_w} F_{\mu\nu}\tilde{Z}^{\mu\nu}\right.
    \nn \\ & ~~~~~~~~~~~~~~\left. + c_{ZZ} \frac{\alpha}{s_w^2 c_w^2} Z_{\mu\nu}\tilde{Z}^{\mu\nu} + c_{WW} \frac{2\alpha}{s_w^2} W_{\mu\nu}^+\tilde{W}^{-\mu\nu} \right), \label{lxi}
\end{align}
where $\sum_\psi$ is over all SM fermions\footnote{We can always ignore the neutrinos because of their small masses.}, $P_L=(1-\gamma^5)/2$, $c_{\gamma\gamma}=c_{WW}+c_{BB}$, $c_{\gamma Z}= c^2_w c_{WW}-s^2_w c_{BB}$, $c_{ZZ}= c^4_w c_{WW}+s^4_w c_{BB}$, and $s_w,c_w$ the sine and cosine of the Weinberg angle.

The fermionic couplings are
\begin{align}
    &\xi_{ee}=\xi_{\mu\mu}=\xi_{\tau\tau}=c_{e_R}-c_{e_L},\nn\\ 
    &\xi_{uu}=\xi_{cc}=c_{u_R}-c_{u_L}, \nn\\ 
    &\xi_{dd}=c_{d_R} - c_{u_L} + (c_{u_L}-c_{t_L})|V_{td}|^2,\nn\\
    &\xi_{ss}=c_{d_R} - c_{u_L} + (c_{u_L}-c_{t_L})|V_{ts}|^2,\nn\\
    &\xi_{bb}=c_{b_R} - c_{u_L} + (c_{u_L}-c_{t_L})|V_{tb}|^2,\nn\\
    &\xi_{tt}=c_{t_R}-c_{t_L},\nn\\
    &\xi_{ds}=\xi_{sd}^*=(c_{t_L}-c_{u_L}) V_{td}^* V_{ts},\nn\\
    &\xi_{db}=\xi_{bd}^*=(c_{t_L}-c_{u_L}) V_{td}^* V_{tb},\nn\\
    &\xi_{sb}=\xi_{bs}^*=(c_{t_L}-c_{u_L}) V_{ts}^* V_{tb}, \label{xiUV}
\end{align}
all evaluated at the UV scale $\Lambda$. For $c_{u_L}\not=c_{t_L}$ there is FV after rotating to the mass eigenbasis, since in general $[V, C_Q]\not =0$. However, the structure of $C_Q$, having two coincident diagonal elements, and the unitarity of the CKM matrix allow one to write the FV couplings as $\xi_{ij} \propto V_{ti}^* V_{tj}$.

In its alternative form (\ref{lxi}) reads,
\begin{align}
    {\mathcal{L}}_{alp} =&-i\frac{a}{f}\sum_{\psi} \xi_{\psi\psi} m_\psi \bar{\psi}\gamma^5\psi \nn\\
    & +i\frac{a}{f}\left(\xi_{ds} m_s \bar{d}P_R s + \xi_{db} m_b \bar{d} P_R b +\xi_{sb} m_b \bar{s} P_R b + \hc\right)\nn\\
    &-i\frac{a}{f}\left(\xi_{ds} m_d \bar{d}P_L s + \xi_{db} m_d \bar{d} P_L b +\xi_{sb} m_s \bar{s} P_L b + \hc\right)\nn\\
    &+ \frac{a}{4\pi f}\left(\tilde c_{GG} \alpha_s G_{\mu\nu}^a\tilde{G}^{a\mu\nu} + \tilde c_{\gamma \gamma} \alpha F_{\mu\nu}\tilde{F}^{\mu\nu} + \tilde c_{\gamma Z} \frac{2\alpha}{s_w c_w} F_{\mu\nu}\tilde{Z}^{\mu\nu}\right.
    \nn \\ & ~~~~~~~~~~~~~~\left. + \tilde c_{ZZ} \frac{\alpha}{s_w^2 c_w^2} Z_{\mu\nu}\tilde{Z}^{\mu\nu} + \tilde c_{WW} \frac{2\alpha}{s_w^2} W_{\mu\nu}^+\tilde{W}^{-\mu\nu} \right), \label{lxialt}
\end{align}
where $\tilde c_{\gamma\gamma}= \tilde c_{WW}+\tilde c_{BB}$, and so on. As a consistency check, $\tilde c_{GG} = c_{GG} + \frac{1}{2}\sum_q \xi_{qq}$ yields the same result as (\ref{CVV}).
Since $m_b\gg m_s\gg m_d$ the FV right handed operators give larger contributions.

\section{The running of the ALP couplings}
\label{sec:running}

The couplings used in Sec.~\ref{sec:ALPlag} are intended at the UV scale $\Lambda\approx 13\mbox{ TeV}$. As such, they are not directly measurable. We need first to run them down to the EW scale, say, just above the top mass $m_t$. Furthermore, for ALPs much lighter than the EW scale, we may also need to integrate out the four heavy SM fields $t, h, Z, W$, match, and run down to the ALP mass $m_a$. The relevant formulas have been collected in \cite{Bauer:2020jbp,Bauer:2021mvw}. We present a streamlined version of their results, keeping only the leading corrections, and refer to the original papers for details. 

Here however, we face a conceptual issue. One difference between the CHM case and the majority of ALPs studies is that in CHM there is often a whole BSM sector just above the EW scale, consisting of additional pNGBs. This precludes the possibility of studying with the same precision, in a model independent way, the RG evolution from the UV scale down to the EW scale. 
Studying the required modifications for each specific model is too involved, given the lack of a clear candidate. 
However,  the couplings between the ALP and other pNGBs stems from the spurion potential that also gives mass
to the latter \cite{Belyaev:2016ftv}. For the ALP to remain light, it should couple only weakly to such potential. In addition, the effect of BSM states below $\Lambda \approx 13\mbox{ TeV}$ on SM higher dimensional operators is already highly constrained. Hence, we assume that the effect of the pNGBs is negligible and estimate the running from the known SM formulas.

Running from $\Lambda$ to $m_t$ does not affect the couplings $c_{GG}, c_{WW}$, and $c_{ZZ}$, and the couplings $\tilde c_{\dots}$ are affected indirectly by the running of the fermionic couplings. Because of the arbitrariness in the remaining fermionic couplings $c_{\dots}$, due to the possibility of shifting them by an overall quantity proportional to baryon number or lepton number, it is better to consider the manifestly invariant values $\xi_{\dots}$. The leading corrections come from the top quark and read \cite{Bauer:2020jbp,Bauer:2021mvw}
\begin{align}
    &\xi_{ee}(m_t)=\xi_{\mu\mu}(m_t)=\xi_{\tau\tau}(m_t)= \xi_{ee}(\Lambda) + 0.116\; \xi_{tt}(\Lambda),\nn\\ 
    &\xi_{uu}(m_t)=\xi_{cc}(m_t)= \xi_{uu}(\Lambda) - 0.116\; \xi_{tt}(\Lambda), \nn\\ 
    &\xi_{dd}(m_t)= \xi_{dd}(\Lambda) + 0.116\; \xi_{tt}(\Lambda) ,\nn\\
    &\xi_{ss}(m_t)= \xi_{ss}(\Lambda) + 0.116\; \xi_{tt}(\Lambda) ,\nn\\
    &\xi_{bb}(m_t)= \xi_{bb}(\Lambda) + 0.097\; \xi_{tt}(\Lambda),\nn\\
    &\xi_{tt}(m_t)= 0.826\; \xi_{tt}(\Lambda),\nn\\
    &\xi_{ds}(m_t) =\xi_{ds}(\Lambda) + 0.0193\;  \xi_{tt}(\Lambda) V_{td}^* V_{ts},\nn\\
    &\xi_{db}(m_t) =\xi_{db}(\Lambda) + 0.0193\; \xi_{tt}(\Lambda) V_{td}^* V_{tb},\nn\\
    &\xi_{sb}(m_t) =\xi_{sb}(\Lambda) + 0.0193\; \xi_{tt}(\Lambda) V_{ts}^* V_{tb}, \label{ximt}
\end{align}
where $\xi_{\dots}(\Lambda)$ are those in (\ref{xiUV}).

Apart from the expected shifts in the diagonal couplings, we see that even if one starts with a flavor preserving Lagrangian in the UV, ($c_{t_L}=c_{u_L}$), running by a couple of orders of magnitude ($\Lambda\to m_t$) already introduces a significant amount of FV. However, notice that the FV is always proportional to the same combination of CKM elements. For instance,
\beq
    \xi_{sb}(m_t) =\left(c_{t_L} -c_{u_L} + 0.0193\;  (c_{t_R} -c_{t_L} )\right) V_{ts}^* V_{tb}. \label{FVatmt}
\eeq

It is in principle possible that the UV coefficients $c_{t_L},c_{u_L}$, and $c_{t_R}$ conspire to make the combination in (\ref{FVatmt}) small. However this would be a highly fine-tuned assumption, devoid of any theoretical justification. The only non-fine-tuned way to have (\ref{FVatmt}) to vanish would be to set $c_{t_L}=c_{u_L}=c_{t_R}$ by some additional symmetry, but this runs against the whole CHM ALP philosophy. Thus, we will see in Sec.~\ref{sec:ALPpheno} that a CHM ALP of mass below that of the $B$ meson is highly disfavored, modulo the above caveats.

If the ALP is much lighter than the EW scale, and we are interested in its on-shell physics, or in its contribution to low energy observables, we need to integrate out the four heavy SM fields and run the obtained coefficients further down to the ALM mass, which we assume to be a few GeVs. We denote the new coefficients by $\eta_{\dots}$ and present their relation to the $\xi_{\dots}$ in (\ref{ximt}), i.e. those evaluated at the EW scale. The leptonic couplings $\eta_{\ell\ell}$ are essentially unaffected. The diagonal quark couplings $\eta_{qq}$, ($q\not = t$), receive a common shift from the running. The FV quarks couplings receive an additional (small) contribution from the matching. 
All in all
\begin{align}
    &\eta_{\ell\ell} = \xi_{\ell\ell},\nn\\
    &\eta_{qq} = \xi_{qq} + (3.0 c_{GG} + 1.5\xi_{uu}+ 1.5\xi_{cc}+ 1.5\xi_{dd}+ 1.5\xi_{ss} +0.9 \xi_{bb})\times 10^{-2},\nn\\
    &\eta_{ds}=\xi_{ds} -V_{td}^* V_{ts} (6.8\, \xi_{tt} + 2.8\, \tilde c_{WW})\times 10^{-5},\nn\\
    &\eta_{db}=\xi_{db} -V_{td}^* V_{tb} (6.8\, \xi_{tt} + 2.8\, \tilde c_{WW})\times 10^{-5},\nn\\
    &\eta_{sb}=\xi_{sb} -V_{ts}^* V_{tb} (6.8\, \xi_{tt} + 2.8\, \tilde c_{WW})\times 10^{-5}. \label{etama}
\end{align} 
Notice that the new FV contributions are also proportional to the same combination of CKM matrix elements, i.e. $\eta_{ij}\propto V_{ti}^* V_{tj}$ as well. 
The diagonal quark couplings are affected at the \% level. Their running in (\ref{etama}) required doing some reverse engineering on the computation of \cite{Bauer:2020jbp,Bauer:2021mvw} since they use  a higher scale as a starting point. ($\xi_{tt}$ is also present, but it is multiplied by the accidentally small coefficient $0.016$.) 

The remaining contributions are flavor diagonal and very small. We do not write them, although they have been computed in \cite{Bauer:2020jbp,Bauer:2021mvw} and can be easily reinstated. To avoid confusion, notice that the much larger top contribution, as well as the contribution $\propto c_{GG}$ quoted in \cite{Bauer:2021mvw}, arise from the RG from the UV scale, and are thus not included in (\ref{etama}). (The top contribution is explicitly indicated in (\ref{ximt}).)

We can now write down the Lagrangian at the ALP mass
\begin{align}
    {\mathcal{L}}_{alp} =& \frac{\partial_\mu a}{2f}\sum_{\psi\not = t} \eta_{\psi\psi} \bar{\psi}\gamma^\mu\gamma^5\psi +\frac{\partial_\mu a}{f}\left(\eta_{ds}\bar{d}\gamma^\mu P_L s + \eta_{db}\bar{d}\gamma^\mu P_L b + \eta_{sb}\bar{s}\gamma^\mu P_L b + \hc\right)\nn\\
    &+ \frac{a}{4\pi f}\left(c_{GG} \alpha_s G_{\mu\nu}^a\tilde{G}^{a\mu\nu} + c_{\gamma \gamma} \alpha F_{\mu\nu}\tilde{F}^{\mu\nu}\right). \label{leta}
\end{align}
At the cost of being pedantic, let us also write the alternative form, analogous to (\ref{lxialt}), since this is the form of the Lagrangian most useful for phenomenological studies.
\begin{align}
    {\mathcal{L}}_{alp} =&-i\frac{a}{f}\sum_{\psi\not=t} \eta_{\psi\psi} m_\psi \bar{\psi}\gamma^5\psi \nn\\
    & +i\frac{a}{f}\left(\eta_{ds} m_s \bar{d}P_R s + \eta_{db} m_b \bar{d} P_R b +\eta_{sb} m_b \bar{s} P_R b + \hc\right)\nn\\
    &-i\frac{a}{f}\left(\eta_{ds} m_d \bar{d}P_L s + \eta_{db} m_d \bar{d} P_L b +\eta_{sb} m_s \bar{s} P_L b + \hc\right)\nn\\
    &+ \frac{a}{4\pi f}\left(\hat c_{GG} \alpha_s G_{\mu\nu}^a\tilde{G}^{a\mu\nu} + \hat c_{\gamma \gamma} \alpha F_{\mu\nu}\tilde{F}^{\mu\nu} \right), \label{letaalt}
\end{align}
where the only point worth stressing is that now\footnote{Awesome notation: $3^{\delta_{q\psi}}=3$ if $\psi$ is a $q$uark, and $=1$ if it is not.}
\begin{align}
    &\hat c_{GG} = c_{GG} + \frac{1}{2}\sum_{q\not= t}\eta_{qq} \approx \tilde c_{GG}-\frac{1}{2}(c_{t_R}-c_{t_L}), \nn\\
    &\hat c_{\gamma\gamma} = c_{\gamma\gamma} + \sum_{\psi\not= t}3^{\delta_{q\psi}} Q_\psi^2 \eta_{\psi\psi}\approx \tilde c_{\gamma\gamma}-\frac{4}{3}(c_{t_R}-c_{t_L}),
\end{align}
the last equalities being satisfied at the \% level\footnote{Eq. (\ref{etama}) can be used to get the exact form.} and $\tilde c_{GG}, \tilde c_{\gamma\gamma}$ are computed by the anomaly of the hyperquark sector.

One could also consider the coupling of very light ALPs to the QCD chiral Lagrangian. This is amply discussed in the literature and does not introduce further issues so we will not pursue it here. Our main interest is in ALPs with a larger mass anyway, since this allows for more relaxed constraints from FV as needed for the generic case $c_{u_L}\not=c_{t_L}$.

\section{The CHM ALP phenomenology}
\label{sec:ALPpheno}

One of the motivations for this work is to further the study \cite{Cacciapaglia:2017iws,Cacciapaglia:2019bqz,Borsato:2021aum,BuarqueFranzosi:2021kky} of an ALP with a mass between the $D^+ D^-$ threshold and the EW scale. We thus distinguish two mass regions:
\begin{itemize}
    \item A light\footnote{Light in this context. Of course, in Dark Matter studies much lighter ALPs are considered.} ALP
    $2 m_D < m_a < m_B$.
    \item A heavy ALP $m_B < m_a < m_W$.
\end{itemize}

Let us begin in the small mass region. Recall the partial decay widths into leptons $\ell$, heavy quarks $q$, photons $\gamma$, and gluons $g$ \cite{Spira:1995rr} (i.e. light hadrons)\footnote{For our simple purpose, we oversimplify a bit the decay width into bosons by using $\hat c_{\gamma\gamma}$ and  $\hat c_{GG}$ instead of the exact one loop result. The expressions below are numerically accurate for $m_a\gg m_b$.}
\begin{align}
    \Gamma(a\to \ell\ell) &=\frac{m_a m_\ell^2}{8\pi f^2}|\eta_{\ell\ell}|^2\sqrt{1-\frac{4m^2_\ell}{m_a^2}}\nn\\
    \Gamma(a\to qq) &=\frac{3 m_a m_q^2}{8\pi f^2}|\eta_{qq}|^2\sqrt{1-\frac{4m^2_{D,(B)}}{m_a^2}}\nn\\
    \Gamma(a\to \gamma\gamma) &=\frac{\alpha^2 m_a^3}{64\pi^3 f^2}|\hat c_{\gamma\gamma}|^2\nn\\
    \Gamma(a\to gg) &=\frac{\alpha_s^2 m_a^3}{8\pi^3 f^2}|\hat c_{GG}|^2\left(1+\frac{83}{4}\frac{\alpha_s}{\pi}\right) \label{branch}
\end{align}
For the smallest value of $m_a$ considered in this work ($m_a = 2 m_D = 3.74\mbox{ GeV}$), and for $\eta_{\ell\ell}=1$, $f=1 \mbox{ TeV}$, we obtain
$\Gamma(a\to \tau\tau)=1.5\times 10^{-7}\mbox{ GeV}$. $\Gamma(a\to c\bar c)$ yields similar results, away from threshold. Thus, generically, the ALP decay is \emph{always prompt}. 

Since in these models the coupling to leptons is ``universal'', (i.e. $\eta_{ee}=\eta_{\mu\mu}=\eta_{\tau\tau}$), we can find an upper bound on the branching ratio of the ALP into electrons and muons by assuming that the total width is  dominated by $a\to\tau\tau$. In this way, all dependence on the ALP couplings to leptons cancels.  
This is not a bad approximation, since the $a\to c\bar c$ channel is closed near threshold, and the di-boson channels are suppressed by numerical factors and powers of the gauge couplings. We find the bounds $BR(ee)\leq 2.66\times 10^{-7}$ and $BR(\mu\mu)\leq 0.0113$, attained at threshold. 

Let us see what these conservative bounds lead to. Table~1 of \cite{Bauer:2021mvw} collects the constraints from rare meson decay. The ones relevant for this case are those involving $B\to\pi\ell\ell$ and  $B\to K \ell\ell$, the strongest one being $B\to K\mu \mu$, yielding\footnote{Strictly speaking, this bound does not extend all the way to $m_B$, since the available phase space closes at $m_a \approx 4.6 \mbox{ GeV}$.}
\begin{align}
    \frac{\eta_{sb}}{V^*_{ts}V_{tb}}\sqrt{BR(\mu\mu)}  <1.1\times 10^{-6}\frac{f}{\mbox{TeV}} \Rightarrow  \frac{\eta_{sb}}{V^*_{ts}V_{tb}} 
   <1.0\times 10^{-5}\frac{f}{\mbox{TeV}}. \label{thebound}
\end{align}
The corrections induced by integrating out the heavy SM fields (\ref{etama}) are of the same order of magnitude, but the running (\ref{FVatmt}) requires the highly fined tuned condition
\begin{align}
   c_{t_L}-c_{u_L} + 0.0193 (c_{t_R} - \, c_{t_L} ) \lesssim 1.0 \times 10^{-5}\frac{f}{\mbox{TeV}}, 
\end{align}
disfavoring the existence of an ALP of this type with a mass $m_a < 4.6\mbox{ GeV}$.

Adding more partial widths to the total width improves the bound, but only slightly, unless $\eta_{\ell\ell}\ll \eta_{qq}$. As an illustration, if we simply take all couplings in (\ref{branch}) to be equal, at threshold (\ref{thebound}) is hardly changed -- simply replace $1.0$ with  $1.1$. 
At the other end of the window ($m_a=4.6\mbox{ GeV}$) we must replace $1.0 \to 1.4$. Thus, we see that (\ref{thebound}) is rather stable against variations of the couplings.

The only radical deviation occurs for $\eta_{\ell\ell} \ll \eta_{qq}$. In this case, away from threshold, the total decay width is dominated by $a\to c\bar c$ and (\ref{thebound}) becomes, for $m_a=4.6\mbox{ GeV}$,
\beq
         \frac{\eta_{sb}}{V^*_{ts}V_{tb}}  \frac{\eta_{\ell\ell}}{\eta_{qq}} <1.3 \times 10^{-5}\frac{f}{\mbox{TeV}}. \label{thenewbound}
\eeq
%In the case where (\ref{thenewbound}) is satisfied, the lack of similar experimental bounds for non-leptonic ALP decay modes would leave the model essentially unconstrained.

\begin{figure}[t]
\begin{center}
\includegraphics[width=0.4\textwidth]{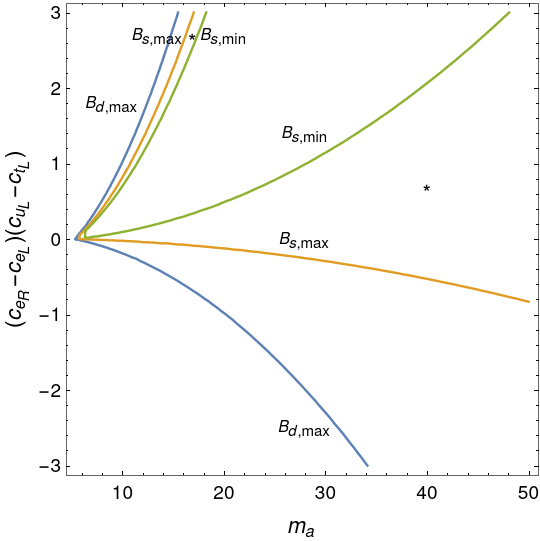}
\caption{Combination of ALP couplings allowed by $B_{d,s}\to\mu\mu$ decay, vs. the ALP mass $m_a$ in GeV.
The blue line, labeled $B_{d,\mathit{max}}$ shows the region where eq.~(\ref{ratioofBR}) reaches the current experimental bound, Similarly, $B_{s,\mathit{min}}$ (green) and $B_{s,\mathit{max}}$ (yellow) denote the two-sigma min. and max. values. The allowed regions, denoted by a $*$, are in the two wedges between the green and yellow lines.}
\label{fig:Bmumu}
\end{center}
\end{figure}

One should also check what is the leading production mode for this ALP. The two contenders are gluon fusion and secondary production via $B$ meson decay. In our previous work \cite{BuarqueFranzosi:2021kky}, we estimated the gluon fusion production cross section for these models to be within 60. and 600.~nb. The total $B\bar B$ production cross section at LHC is of the order of $5.\times 10^5\mbox{ nb}$. In order to be competitive with gluon fusion, it requires a branching ratio $B\to a X$ between $10^{-3}$ and $10^{-4}$. 

We assume $|\eta_{sb}| = |V^*_{ts}V_{tb}|\times 10^{-5}$ which is the maximum amount allowed by FV for $f=1 \mbox{ TeV}$.
Using for simplicity the quark level formula
\begin{align}
    \Gamma(b\to a s) = \frac{1}{32 \pi f^2}|\eta_{sb}|^2 m_b^3 \left(1-\frac{m_a^2}{m_B^2}\right)^2
\end{align}
yields, at threshold ($m_a=2 m_D$), $\Gamma(b\to a s) =4.5 \times 10^{-20}\mbox{ GeV}$, corresponding to a branching ratio of $1.1\times 10^{-7}$, well below the required value. Thus, for the present models, gluon fusion is the main production mode. It is only for $\hat c_{GG}/f\ll1/\mbox{TeV}$ that the $B \bar B$ production mode becomes competitive as in e.g. \cite{DallaValleGarcia:2023xhh}.

Let us conclude by moving into the high mass region and study what regions of parameter space are allowed by $B_{d,s}\to\mu\mu$ decay via an off-shell ALP.
The comparison between the branching ratio induced by the ALP (including interference) and that of the SM is also given in 
\cite{Bauer:2021mvw} (see also \cite{Hiller:2014yaa}) and reads, for the models at hand
\begin{align}
    BR_{a+SM}=BR_{SM}\left|1+\frac{(c_{e_R}-c_{e_L})(c_{u_L}-c_{t_L})}{4.2}\frac{\pi}{\alpha}\frac{v^2}{f^2}\frac{1}{1-m^2_a/m^2_B}\right|^2. \label{ratioofBR}
\end{align}

Taking the SM central values computed in \cite{Beneke:2019slt} $BR_{SM}(B_d\to\mu\mu)=1.03\times 10^{-10}$ and $BR_{SM}(B_s\to\mu\mu)=3.66\times 10^{-9}$,  and the experimental values \cite{ParticleDataGroup:2024cfk} $BR_{exp}(B_d\to\mu\mu)<1.5\times 10^{-10}$ and $BR_{exp}(B_s\to\mu\mu)=(3.34\pm 0.27)\times 10^{-9}$, allows to set bounds in the $m_a$ vs. $(c_{e_R}-c_{e_L})(c_{u_L}-c_{t_L})$ plane, shown in Fig.~\ref{fig:Bmumu}. We see that, for $m_a\gg m_B$, FV terms $c_{u_L}-c_{t_L}\approx 1$ are in principle allowed in this mass range.

\section*{Acknowledgments}
I would like to thank Diogo Buarque Franzosi, Giacomo Cacciapaglia, Xabier Cid Vidal, Thomas Flacke, Carlos Vázquez Sierra, and Hugo Serôdio for our collaborations in the study of the composite ALP. Thanks to the organizers of the workshop FIPs@LHCb at CERN and to Maksym Ovchynnikov and Giovani Dalla Valle Garcia for valuable comments that led to some of the issues discussed here. I also wish to thank SISSA for the hospitality during the completion of this work, and David Marzocca for enlightening discussions while there.

I am supported by the Swedish Research Council (grant nr. 2024-04347), as well as by grants from the Adlerbert Research Foundation via the KVVS foundation, the Carl Tryggers Foundation (CTS 24:3453) and the Kungl. Vetenskapsakademien (PH2024-0076).

% % % % % % % % % % % % % % % % % % % % % % % % % % % % % % % % % % % % % %
\bibliography{main.bib}
\bibliographystyle{JHEP}
% % % % % % % % % % % % % % % % % % % % % % % % % % % % % % % % % % % % % %

\end{document}